# Blind Denoising Autoencoder

Angshul Majumdar

*Abstract*—The term 'blind denoising' refers to the fact that the basis used for denoising is learnt from the noisy sample itself during denoising. Dictionary learning and transform learning based formulations for blind denoising are well known. But there has been no autoencoder based solution for the said blind denoising approach. So far autoencoder based denoising formulations have learnt the model on a separate training data and have used the learnt model to denoise test samples. Such a methodology fails when the test image (to denoise) is not of the same kind as the models learnt with. This will be first work, where we learn the autoencoder from the noisy sample while denoising. Experimental results show that our proposed method performs better than dictionary learning (K-SVD), transform learning, sparse stacked denoising autoencoder and the gold standard BM3D algorithm.

## I. Introduction

THE simplest way to denoise a natural signal is to assume that the signal is sparse in some transform (wavelet, DCT, Gabor etc.) whereas the noise is not. Given the statistical properties of the noise, one can threshold in the transform domain to get rid of the noise and preserve only the high valued signal coefficients. A clean version of the signal is reconstructed by applying the inverse transform. Signal and image processing literature has seen many variants of this basic idea [1-3].

However the problem with this approach is that the denoising performance is limited by the sparsifying capacity of the transform – sparser the representation better the denoising. But one does not know what is the best basis for representing the particular sample that needs to be denoised – is it wavelet, is it DCT or something more sophisticated like curvelet or contourlet. It is a chicken and egg problem, to know the best sparsifying basis one needs access to the clean image, but that (denoising) is the problem one needs to solve.

The limitation of fixed transform paved way for learning based denoising. Few of them are based on the kernel PCA technique [4, 5]. The more popular and successful ones are based on adaptive learning of the sparsifying basis. K-SVD [6, 7] is perhaps the most popular technique; it is based on the dictionary learning approach. Another (more recent) technique is based on transform learning [8, 9]. Both of them are blind denoising techniques, i.e. they learn the sparsity basis adaptively from the signal while denoising.

A. Majumdar is with Indraprastha Institute of Information Technology, New Delhi, India. (e-mail: angshul@iiitd.ac.in).

In recent times a handful of papers have been published on autoencoders used for signal denoising [10-13]. These techniques learn the autoencoder model on a training set and uses the trained model to denoise new test samples. Unlike the dictionary learning and transform learning based approaches, the autoencoder based methods were not blind; they were not learning the model from the signal at hand.

The main disadvantage of such an approach is that, one never knows how good the learned autoencoder will generalize on unseen data. In the aforesaid studies [10-13] training was performed on standard dataset of natural images (image-net / CIFAR) and used to denoise natural images. Can the learnt model be used to recover images from other modalities – radar, SAR, MRI, CT etc.? In this work we will show that the answer is in the negative. These techniques can only denoise when large volume of training data is available from the same modality. This is not a problem for natural images – however the usefulness of denoising natural images is questionable. Usually digital photographs are very clean, one does not need denoising; it is only for scientific imaging modalities (mentioned before) denoising becomes an important pre-processing step. Unfortunately for such modalities, large volume of training data is not readily available. Therefore in such cases the autoencoder based approaches are likely to fail; models learnt on natural images do not generalize well on other modalities.

This work proposes an autoencoder based formulation for blind denoising; i.e. we do not need to train the autoencoder to denoise on a separate training dataset. It will learn the autoencoder from the signal while denoising. The proposed approach yields better results than K-SVD, transform learning and prior autoencoder based approaches. In fact it yields even better results than the gold standard BM3D algorithm on an average.

## II. Background

### A. Dictionary Learning based Denoising

Dictionary learning is a synthesis formulation; it learns a dictionary (D) so as to synthesize / re-generate the data (X) from the learned coefficients (Z).

$$X = DZ \qquad (1)$$

Basically it factorizes the data matrix into a dictionary and coefficients. The topic has been around from the late 90s [14, 15]. However the term 'dictionary learning' is recent. In early days, the technique was used to learn filters mimicking early human vision. The solution to (1) is formulated as follows,

$$\min_{D,Z} \|X - DZ\|_F^2 \text{ such that } \|Z\|_0 \leq \tau \qquad (2)$$

Here the l0-norm (defined as the number of non-zero elements) is defined on the vectorized version of the matrix. The parameter $\tau$ controls the sparsity level. The first term enforces data fidelity and the constraint promotes sparsity in the coefficients. It (2) is solved via alternative minimization. In one step, the dictionary / codebook is updated assuming the coefficients to be fixed and in the next the sparse-code / coefficients are updated assuming the dictionary to be given. There are many sophisticated algorithms to solve such matrix factorization problems [16, 17]; but for all practical cases a simple alternating least squares works well.

Today dictionary learning has a widespread application in image processing and computer vision. However vision is not the area of our interest. We will restrict ourselves to its application in inverse problems – mainly on denoising.

A linear inverse problem can be expressed in the form –

$$y = Ax \quad (3)$$

where x is to be solved, y is the measurement and A the linear operator.

Dictionary learning is a patch based formulation. The dictionary is learnt from patches of the signal. This is expressed as,

$$P_i \hat{x} = D z_i \ \forall i \quad (4)$$

Here $P_i$ is the patch extraction operator. This expression is equivalent to (1). Here $\hat{x}$ is an estimate of the signal to be solved.

Solution of the linear inverse problem via dictionary learning proceeds in two broad phases. At first an approximate solution to (3) is obtained using the dictionary and its coefficients; next this approximate solution is used in dictionary learning. The entire formulation is given by,

$$\min_{\hat{x}, D, Z} \|y - A\hat{x}\|_2^2 + \lambda \left( \sum_i \|P_i \hat{x} - D z_i\|_F^2 \ \text{s.t.} \ \|z_i\|_0 \leq \tau \right) \quad (5)$$

Here Z is formed by stacking the patches $z_i$ as columns.

Depending on the nature of the linear operator, various kinds of inverse problems can be solved. For denoising [7] A is identity. For inpainting [18] A is a restriction operator; for reconstruction [19] it is a projection.

For denoising, the Euclidean norm cost function (5) is suitable for the case of Gaussian noise. For other types of noise, the cost function needs to be changed accordingly. For example in impulse denoising the data fidelity term needs to be changed to absolute deviations [20], leading to the following formulation.

$$\min_{\hat{x}, D, Z} \|x - \hat{x}\|_1 + \lambda \left( \min_{D, Z} \sum_i \|P_i \hat{x} - D z_i\|_F^2 \ \text{s.t.} \ \|z_i\|_0 \leq \tau \right) \quad (6)$$

Studies like [21] proposed changing the data fidelity term in the dictionary learning to $l_1$-norm; but this is unnecessary; as has been shown in [20]. However both [20] and [21] yield very similar results.

### B. Transform Learning based Denoising

Transform learning is the analysis equivalent of dictionary learning. It analyses the data by learning a transform / basis to produce coefficients. Mathematically this is expressed as,

$$TX = Z \quad (7)$$

Here T is the transform, X the data and Z the corresponding coefficients. The following formulation was proposed [22, 23] –

$$\min_{T,Z} \|TX - Z\|_F^2 + \lambda \left( \|T\|_F^2 - \log \det T \right) \ \text{s.t.} \ \|Z\|_0 \leq \tau \quad (8)$$

The factor $-\log \det T$ imposes a full rank on the learned transform; this prevents the degenerate solution (T=0, Z=0). The additional penalty $\|T\|_F^2$ is to balance scale.

An alternating minimization approach has been proposed to solve the transform learning problem.

$$Z \leftarrow \min_Z \|TX - Z\|_F^2 \ \text{s.t.} \ \|Z\|_0 \leq \tau \quad (9a)$$

$$T \leftarrow \min_T \|TX - Z\|_F^2 + \lambda \left( \|T\|_F^2 - \log \det T \right) \quad (9b)$$

Updating the coefficients (9a) is straightforward; it is a standard sparse coding step. Although solving (9b) seems tricky a closed form update was derived in [23]; given by,

$$XX^T + \lambda \varepsilon I = LL^T \quad (10a)$$

$$L^{-1} XZ^T = USV^T \quad (10b)$$

$$T = 0.5 R \left( S + (S^2 + 2\lambda I)^{1/2} \right) Q^T L^{-1} \quad (10c)$$

In a manner similar to dictionary learning transform learning has been used to solve inverse problems. The general formulation is given by

$$\min_{\hat{x}, T, Z} \|y - A\hat{x}\|_2^2 + \lambda \left( \sum_i \|T P_i \hat{x} - z_i\|_F^2 \ \text{s.t.} \ \|z_i\|_0 \leq \tau \right) \quad (11)$$

One notices that the first term remains as before – this is the data fidelity term. The term within (.) is the transform learning formulation – it is equivalent to (8). The solution to (11) proceeds in two steps. At first the signal is updated by solving,

$$\hat{x} \leftarrow \min_{\hat{x}} \|y - A\hat{x}\|_2^2 + \lambda \sum_i \|T P_i \hat{x} - z_i\|_F^2 \quad (12)$$

This is a simple least square update having a closed form solution. The second step is transform learning, given by –

$$\min_{T,Z} \sum_i \|T P_i \hat{x} - z_i\|_F^2 \ \text{s.t.} \ \|z_i\|_0 \leq \tau \quad (13)$$

This remains the same as (8).

Transform learning is a new approach (less than 5 years old) and is not as popular as dictionary learning. Hence its application has been limited. It has been used in denoising [9] and reconstruction [8]; a comprehensive list on theorey, algorithms and applications of this technique is at [24].

Note that as in dictionary learning, only the data fidelity term in (11) needs to be changed based on the noise model; there is no need to change the cost function for transform learning.

### C. Autoencoder based Denoising

Autoencoders are self-superivsed neural networks. Ideally the input and the output are supposed to be the same. However, it has been found that if the input to the autoencoder is noisy and the output is clean, the autoencoder weights are more robust. Such denoising autoencoders (rather their stacked sparse versions) can be used to clean noisy inputs [10-12].

However stacked sparse denoising autoencoder (SSDA) based denoising is inductive, unlike dictionary and transform learning based techniques, which are transductive in nature. During training the SSDA learns to denoise from large amount of training data; the input to the SSDA are noisy samples and the outputs the corresponding clean samples. Thus it learns to denoise. During testing, the noisy sample is presented at the input and a clean sample is expected at the output of the SSDA.

One advantage that has been empirically claimed is that, one does not need to change the autoencoder training algorithm in any fashion depending on the noise model. Only the training data needs to be corrupted by the type of noise that it needs to clean. However, this is not a particularly elegant approach. Even for one type of noise, different SSDAs need to be trained for different amounts of noise; i.e. an autoencoder learnt to denoise Gaussian noise with 0 mean and standard deviation 25 will not be able to clean noise of 0 mean and standard deviation 5. Two different SSDAs need to be learnt. The same applies for other kinds of noise like impulse, speckle etc.

Blind denoising techniques are more elegant in this respect. There is no training required. One only needs to know the value of few parameters ($\lambda$ / $\tau$) in (5) / (11) required for denoising a particular amount of noise. This can be easily precomputed on a single validation image.

The other problem of SSDA based denoising is that it is heavily dependent on training data; it expects the test data to be similar to the training data. The experiments in the aforesaid papers [10-12] have been carried out on natural images. There is no dearth of natural images on the web, therefore making a large training set is feasible. However, natural images are hardly corrupted by noise in practice. It is the scientific images, like SAR, satellite, MRI, CT, USG modalities that need to be denoised in reality.

As will be shown later, SSDA trained on natural images give poor performance on such scientific imaging modalities. This is because the structure of natural images are significantly different from these. Hence the autoencoder fails to generalize on the unseen modality. Proponents of deep learning would argue that fine-tuning would improve the result. In practice even for fine-tuning a significant volume of data is required. For the said imaging modalities, acquiring such a volume of data is not easy. In most cases the data is proprietary and not publicly available.

III. PROPOSED FORMULATION

A. Gaussian Denoising

This work concentrates on the additive noise model. This is expressed as,

$$x = x_0 + n \tag{14}$$

Here $x_0$ is the clean signal that needs to be recovered; n is the additive noise and x the corrupted noisy signal.

We propose a blind denoising approach, i.e. we will learn an autoencoder while denoising. We do not need access to any training data as required by [10-12].

We learn the autoencoder from patches of the signal. This is expressed as,

$$P_i \hat{x} = W' \varphi(W P_i \hat{x}), \forall i \tag{15}$$

Here (and everywhere else) $\varphi$ denotes the activation function. The learning is based on the usual autoencoder formulation that minimizes the Euclidean cost.

$$\min_{W',W} \sum_i \|P_i \hat{x} - W' \varphi(W P_i \hat{x})\|_2^2 \tag{16}$$

As in other patch-based denoising techniques (5)/(11) we need a global consistency term between the noisy image (x) and the denoised estimate ($\hat{x}$). Assuming Gaussian noise, this will be the simple Euclidean norm: $\|x - \hat{x}\|_2^2$. Therefore the complete denoising formulation takes the form:

$$\min_{W',W,\hat{x}} \|x - \hat{x}\|_2^2 + \lambda \sum_i \|P_i \hat{x} - W' \varphi(W P_i \hat{x})\|_2^2 \tag{17}$$

In all prior studies in denoising, it has been observed that sparsity on the features improves performance [6-12]. Therefore, we incorporate sparsity into (17) by adding an $l_1$-norm penalty on the coefficients.

$$\min_{W',W,\hat{x}} \|x - \hat{x}\|_2^2 \\ + \lambda \sum_i \left( \|P_i \hat{x} - W' \varphi(W P_i \hat{x})\|_2^2 + \mu \|\varphi(W P_i \hat{x})\|_1 \right) \tag{18}$$

This is not an easy problem to solve. We resort to the Split Bregman approach [25-27]. We introduce a proxy variable $z_i = \varphi(W P_i \hat{x})$. After relaxing the equality constraint via the augmented Lagrangian and introducing the Bregman variable our formulation takes the form,

$$\min_{W',W,\hat{x},Z} \|x - \hat{x}\|_2^2 + \lambda \sum_i \left( \|P_i \hat{x} - W' z_i\|_2^2 + \mu \|z_i\|_1 \right) \\ + \gamma \sum_i \|z_i - \varphi(W P_i \hat{x}) - b_i\|_2^2 \tag{19}$$

Here $b_i$'s are the Bregman relaxation variables that are updated automatically in every iteration so as to enforce equality between the variables and their proxy at convergence.

Using alternating direction method of multipliers [28, 29], (19) can be segregated into the following problems.

$$P1: \min_{W'} \sum_i \left( \|P_i \hat{x} - W' z_i\|_2^2 \right)$$

$$P2: \min_W \sum_i \|z_i - \varphi(W P_i \hat{x}) - b_i\|_2^2 \equiv \min_W \sum_i \|\varphi^{-1}(z_i - b_i) - W P_i \hat{x}\|_2^2$$

$$P3: \min_{\hat{x}} \|x - \hat{x}\|_2^2 + \lambda \sum_i \|P_i \hat{x} - W' z_i\|_2^2 + \gamma \sum_i \|z_i - \varphi(W P_i \hat{x}) - b_i\|_2^2$$

$$\equiv \min_{\hat{x}} \|x - \hat{x}\|_2^2 + \lambda \sum_i \|P_i \hat{x} - W' z_i\|_2^2 + \gamma \sum_i \|\varphi^{-1}(z_i - b_i) - W P_i \hat{x}\|_2^2$$

$$P4: \min_Z \lambda \sum_i \left( \|P_i \hat{x} - W' z_i\|_2^2 + \mu \|z_i\|_1 \right) + \gamma \sum_i \|z_i - \varphi(W P_i \hat{x}) - b_i\|_2^2$$

$$\equiv \lambda \sum_i \|P_i \hat{x} - W' z_i\|_2^2 + \mu \|z_i\|_1 + \gamma \|z_i - \varphi(W P_i \hat{x}) - b_i\|_2^2$$

Sub-problem P1 is a simple least squares problem. Sub-problem P2 can be equivalently represented as a least squares problem; this is possible since the activation functions are applied element-wise and hence trivial to invert. Using the same logic P3 can also be recast as a least squares problem. All the least squares problem have analytic solutions in the form of pseudoinverse.

Sub-problem P4 is easily decoupled into solving each of the $z_i$'s separately. This leads to an $l_1$-regularized least squares problem. This is given by –

$$\min_{z_i} \|P_i\hat{x} - W'z_i\|_2^2 + \gamma \|z_i - \varphi(WP_i\hat{x}) - b_i\|_2^2 + \mu \|z_i\|_1$$

It can be easily solved using iterative soft thresholding algorithm [30].

In the final step is to update the Bregman relaxation variables. This is done by simple gradient descent.

$$b_i \leftarrow z_i - \varphi(WP_i\hat{x}) - b_i, \forall i$$

In each iteration we have to solve four sub-problems P1 to P4. P1-P3 are simple linear least squares problems. They have a closed form solution in the form of pseudoinverse. The complexity of computing the pseudoinverse is $O(n^w)$ where w<2.37; this is proven infimum, in practice it is conjectured to be w=2. For solving sub-problem P4, one needs to iterative solve the sparse coding problem. It is usually run for a fixed number of iterations (say k). Each iteration requires two matrix products and one thresholding. The complexity of the matrix products is also $O(n^w)$ and that of thresholding is $O(n)$. Therefore the overall complexity per iteration of the algorithm is $3 \times O(n^w) + k\{O(n^w) + O(n)\}$. The computational complexity of dictionary and transform learning would be $3 \times O(n^3) + k\{O(n^w) + O(n)\}$; it is slightly higher than our proposed technique owing to the requirement of computing singular value decompositions in each iteration.

*B. Impulse Denoising*

So far we have discussed techniques for Gaussian denoising. For impulse denoising, the only change will be in the global data fidelity term of (17); instead of the Euclidean norm we need to minimize the taxi-cab distance.

$$\min_{W',W,\hat{x}} \|x - \hat{x}\|_1 + \lambda \sum_i \left( \|P_i\hat{x} - W'\varphi(WP_i\hat{x})\|_2^2 + \mu \|\varphi(WP_i\hat{x})\|_1 \right) \quad (20)$$

With the same substitution as before $z_i = \varphi(WP_i\hat{x})$, (19) can be expressed as,

$$\min_{W',W,\hat{x},Z} \|x - \hat{x}\|_1 + \lambda \sum_i \left( \|P_i\hat{x} - W'z_i\|_2^2 + \mu \|z_i\|_1 \right) + \gamma \sum_i \|z_i - \varphi(WP_i\hat{x}) - b_i\|_2^2 \quad (21)$$

The sub-problems for (21) will remain almost the same except for the update of $\hat{x}$; instead of P3, we will have –

$$\min_{\hat{x}} \|x - \hat{x}\|_1 + \lambda \sum_i \|P_i\hat{x} - W'z_i\|_2^2 + \gamma \sum_i \|\varphi^{-1}(z_i - b_i) - WP_i\hat{x}\|_2^2 \quad (22)$$

To solve (22), one needs to substitute $y = x - \hat{x}$. This leads to the following in the Split Bregman framework,

$$\min_{\hat{x},y} \|y\|_1 + \varepsilon \|y - x + \hat{x} - c\|_2^2 \\ \lambda \sum_i \|P_i\hat{x} - W'z_i\|_2^2 + \gamma \sum_i \|\varphi^{-1}(z_i - b_i) - WP_i\hat{x}\|_2^2 \quad (23)$$

Here c is the relaxation variable. The variable $\hat{x}$ and its proxy y can be updated in closed forms,

$$y \leftarrow \min_y \|y\|_1 + \varepsilon \|y - x + \hat{x} - c\|_2^2 \quad (24a)$$

$$\min_{\hat{x}} \varepsilon \|y - x + \hat{x} - c\|_2^2 \\ \lambda \sum_i \|P_i\hat{x} - W'z_i\|_2^2 + \gamma \sum_i \|\varphi^{-1}(z_i - b_i) - WP_i\hat{x}\|_2^2 \quad (24b)$$

(23a) can be solved using one step of soft thresholding [1], and (23b) being a least squares problem can be solved analytically via pseudoinverse.

For both impulse and Gaussian denoising the stopping criteria remain the same. We specify a maximum number of iterations (40). The other stopping criterion is local convergence, i.e. when the cost function does not change significantly in subsequent iterations.

In terms of complexity the only change that happens for this problem is in the P3. Earlier it had a closed form solution. Here it needs to be updated iterative via sparse coding. We have already discussed the complexity of the sparse coding step.

## IV. EXPERIMENTAL RESULTS

In this work we compare with dictionary and transform learning based adaptive / blind denoising techniques. We also compare with SSDAs since they have been used in the past for non-blind denoising [10-12]. Recently CNN based non-blind / non-adaptive denoising techniques are also becoming popular, so we compare with one such formulation [13]. As a benchmark we use the gold standard BM3D denoising algorithm.

We carry out experiments on both natural images like Lena, Barbara, Cameraman and Peppers. However, all of them are natural images. We also carry out experiments on MRI (brain and phantom) and hyperspectral image (WDC – Washington DC and Gulf of Mexico). In this work we compare with all the standard denoising approaches – K-SVD [7], Transform [31], SSDA [10], and BM3D. We also compare with the latest deep CNN (DnCNN) denoising method [34]. Given the limitations of this paper, we are unable to repeat the configuration for each of the denoising tools. We request the reader to peruse the said references.

Experiments are conducted at two noise Gaussian noise levels – low (σ=10) and high (σ=100). Peak Signal to Noise Ratio (PSNR) is used as the metric for comparison. The experimental results are shown in Table I.

Our proposed method requires specifying two parameters λ and μ, and one hyper-parameter γ. All the parameters have been tuned on a separate validation image (baboon) via grid search; the values are λ = 0.5, μ=0.1 and γ=0.5. For our proposed method, we have used an autoencoder where the number of nodes in the representation layer are twice (128) the number in the input layer (overlapping patches of 8x8). The encoder layer has been initialized with concatenated wavelet and DCT while the decoder has been initialized with the corresponding inverse transforms.

We have also carried out experiments on impulse denoising. The dictionary learning based method used for impulse denoising is [21]. The transform learning based formulation used for impulse denoising is [32]. The BM3D based formulation used for impulse denoising is [33]. SSDA and CNN based techniques do not need any change any formulation; they only need to be trained on data corrupted by impulse noise. The results are shown in Table II. Experiments are carried out at two noise levels (5% salt and pepper noise and 50% salt and pepper noise).

TABLE I  COMPARATIVE GAUSSIAN DENOISING PERFORMANCE

| Dataset | Low – σ=10 | | | | | | High – σ=50 | | | | | |
|---|---|---|---|---|---|---|---|---|---|---|---|---|
| | KSVD | Transform | SSDA | BM3D | DnCNN | Proposed | KSVD | Transform | SSDA | BM3D | DnCNN | Proposed |
| Lena | 36.91 | 37.62 | 37.90 | 37.92 | **38.02** | 37.66 | 23.49 | 23.66 | 24.24 | 24.02 | 24.21 | **24.87** |
| Barbara | 34.42 | 34.55 | 34.67 | 34.21 | 33.77 | **34.72** | 21.86 | 22.42 | 23.22 | 24.37 | 24.56 | **25.01** |
| Cameraman | 33.72 | 33.87 | 34.26 | 33.98 | 34.14 | **34.68** | 21.75 | 22.01 | 21.98 | 22.42 | 22.55 | **23.02** |
| Peppers | 34.67 | 35.19 | 36.02 | 35.86 | **36.29** | 35.75 | 21.87 | 22.36 | 22.24 | 22.19 | 22.52 | **22.92** |
| Brain | 38.54 | 38.78 | 30.42 | 38.79 | 30.78 | **38.96** | 24.73 | 24.83 | 19.46 | 24.97 | 19.97 | **25.42** |
| Phantom | 36.72 | 36.41 | 29.92 | 36.79 | 30.81 | **36.93** | 23.97 | **24.36** | 19.45 | 24.25 | 19.88 | **24.36** |
| WDC | 33.48 | 33.67 | 28.77 | 33.51 | 30.45 | **34.02** | 22.58 | 22.60 | 17.81 | 22.59 | 18.46 | **23.16** |
| Gulf | 32.19 | 32.41 | 27.92 | 32.87 | 29.73 | **33.00** | 21.86 | 22.33 | 17.64 | 22.12 | 18.02 | **22.73** |

TABLE II  COMPARATIVE IMPULSE DENOISING PERFORMANCE

| Dataset | Low – 5% | | | | | | High – 50% | | | | | |
|---|---|---|---|---|---|---|---|---|---|---|---|---|
| | [21] | [32] | SSDA | [33] | DnCNN | Proposed | [21] | [32] | SSDA | [33] | DnCNN | Proposed |
| Lena | 35.92 | 35.96 | 36.50 | 36.13 | 35.61 | **36.27** | 24.16 | 24.29 | 24.03 | 23.56 | 23.17 | **24.59** |
| Barbara | 32.47 | 33.81 | 33.69 | 33.87 | 33.62 | **34.03** | 23.02 | 23.14 | 22.29 | 21.96 | 21.53 | **23.66** |
| Cameraman | 32.96 | 33.21 | 33.24 | 33.02 | 33.17 | **33.64** | 22.51 | 22.42 | 21.37 | 21.50 | 21.63 | **23.01** |
| Peppers | 34.51 | 34.86 | 35.20 | **35.51** | 34.73 | 35.37 | 22.11 | 22.42 | 22.06 | 21.77 | 21.46 | **22.76** |
| Brain | 35.44 | 35.69 | 28.46 | 34.49 | 28.50 | **35.98** | 23.03 | 23.11 | 18.73 | 22.56 | 18.91 | **23.89** |
| Phantom | 33.85 | 34.11 | 28.29 | 33.56 | 28.62 | **34.43** | 22.22 | 22.53 | 18.62 | 22.19 | 18.88 | **22.95** |
| WDC | 33.19 | 33.52 | 27.03 | 33.26 | 27.31 | **34.01** | 21.72 | 21.96 | 17.16 | 21.11 | 17.56 | **22.45** |
| Gulf | 32.74 | 33.19 | 26.63 | 32.79 | 27.37 | **33.36** | 21.46 | 21.84 | 17.09 | 20.66 | 17.73 | **22.07** |

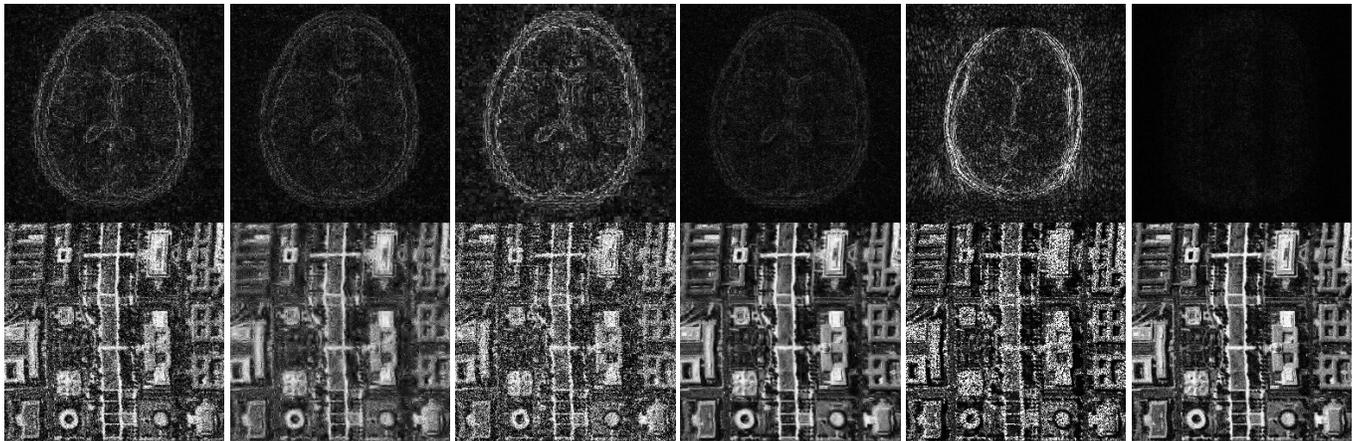

**Fig. 1.** Top. Denoised Difference Images of Brain MRI. Bottom. Denoised Hyperspectral image (band 50) WDC. Left to Right – KSVD, Transform Learning, SSDA, BM3D, DnCNN and Proposed

Experimental results in Tables I and II, show that our proposed method yields the best results for heavy noise. For light noise, our method yields better results than the others in most cases. Especially for scientific imaging modalities (MRI and hyper-spectral) we always yield the best results.

What is interesting to note is that SSDA and CNN yields results at par or better than existing state-of-the-art algorithms only for natural images; but yields results significantly subpar than others for MRI and hyper-spectral denoising. This was expected. Since SSDA and CNN has been trained on a large volume of natural images, its performance on similar images are good, but fails to generalize on unseen imaging modalities.

For visual evaluation we show results on the brain MRI and the one spectral band (number 50) of the hyperspectral image of WDC.

For the brain, we follow the convention of medical imaging. We show the difference (between denoised and original) image. This helps in visually understanding the denoising artifacts. These are shown in Fig. 1. For visual clarity, the images have been contrast enhanced 10 times. Owing to limitations in space, we are only showing results for high Gaussian noise (σ=50). The results corroborate the numerical metrices. Difference image from our proposed method is almost completely dark, meaning that there are hardly any denoising artifacts. The artifacts are slightly more pronounced in BM3D, Transform learning and KSVD; but are the worst in SSDA.

For the hyperspectral image, we show denoising results for low Gaussian noise (σ=10). The conclusions remain the same. KSVD is a bit noisy but maintains detailed edges. The transform learning approach overtly smooths the image. But the worst one is from SSDA. BM3D yields good results; but ours are better. The sharpness is better preserved.

In this work, we do not tabulate the actual run-times. On an average for Gaussian denoising of 256 x 256 images (Lena, Barbara, Cameraman, Peppers, Brain and Phantom) DL takes about 15 seconds and TL about 16 seconds. Our proposed method is slightly faster and takes 13 seconds. The SSDA is very fast and takes only 0.13 seconds, the CNN takes about 0.76 seconds. All the experiments have been carried out on an intel i7 PC with 16 GB of RAM running MATLAB R2012a.

## V. CONCLUSION

This work introduces a new adaptive denoising technique based on autoencoders. Unlike prior studies [10-13] that require huge volume of training data for learning the autoencoder denoising model, our approach is completely blind. It does not require any training data. It learns the autoencoder model while denoising.

The reason why our method yield better results compared to dictionary and transform learning based techniques can be understood thus. Consider an image of size 256x256. Assuming a square dictionary and transform with 8x8 patches we need to learn a dictionary / transform of size 64x64 and coefficients of size 64x1024. For our formulation one only need to learn encoders and decoders of sizes 64x64. Thus the number of parameters we need to estimate are an order of magnitude less compare to existing techniques.

Experiments have been carried out on a variety of images. Comparison has been done with all well known denoising techniques – KSVD, Transform learning, SSDA, CNN and BM3D. Overall, we always perform the best.

## ACKNOWLEDGEMENT

This work is supported by DST-CNRS-2016-02.